\pdfoutput=1
\documentclass[aps,pra,twocolumn,superscriptaddress,nofootinbib,a4paper]{revtex4-2}
\usepackage{amsmath}
\usepackage{booktabs}
\usepackage{amssymb}
\usepackage{graphicx}
\usepackage{bbm}
\usepackage{tabularx}
\usepackage{xcolor}
\usepackage{multirow}
\definecolor{myblue}{RGB}{43,140,190}
\definecolor{mybluemid}{RGB}{166,189,219}
\definecolor{mybluelight}{RGB}{236,231,242}
\newcommand{\asym}[3]{$#1^{+#2}_{-#3}$}

\usepackage{array}
\usepackage{booktabs}
\usepackage{tikz}
\usetikzlibrary{calc, matrix, positioning, arrows.meta}
\usepackage{placeins}
\usepackage{float}
\usepackage{hyperref}

\makeatletter
\def\keeparray@list{\@array\@@array\@tabular\@tabarray\array\endarray\endtabular\@mkpream\@classx\insert@column\@arraycr\@xarraycr\@xargarraycr\@yargarraycr}
\def\keeparray@save#1{\ifx#1\@nil\else\expandafter\let\csname keeparray@\string#1\endcsname#1\expandafter\keeparray@save\fi}
\def\keeparray@load#1{\ifx#1\@nil\else\expandafter\let\expandafter#1\csname keeparray@\string#1\endcsname\expandafter\keeparray@load\fi}
\expandafter\keeparray@save\keeparray@list\@nil
\AtBeginDocument{\expandafter\keeparray@load\keeparray@list\@nil
  \expandafter\let\csname endtabular*\endcsname\endtabular}
\makeatother

\newcommand{\imperial}{Department of Computing, Imperial College London, 180 Queen's Gate, SW7 2AZ, UK}
\newcommand{\edinburgh}{School of Informatics, The University of Edinburgh, 10 Crichton St, EH8 9AB, UK}

\begin{document}

\title{Graph Neural Post-selection for Quantum Error Correction}

\author{Conor Carty}
\email{c.carty24@imperial.ac.uk}
\affiliation{\imperial}
\author{Tamas Noszko}
\affiliation{\edinburgh}
\author{Joschka Roffe}
\affiliation{\edinburgh}
\author{Roberto Bondesan}
\affiliation{\imperial}

\begin{abstract}
Post-selection improves the logical reliability of quantum computation by rejecting shots which are likely to result in logical failure. We introduce graph neural networks that predict decoder failure without additional decoder executions, using only syndromes, and for high-rate quantum low-density parity-check (qLDPC) codes, existing belief-propagation posteriors. We evaluate rotated surface and bivariate bicycle (BB) code memories under uniform depolarising circuit-level noise. Retaining 90\% of shots gives approximately $3700\times$ and $740 \times$ reductions in logical error rates for $[[72,12,6]]$ and $[[144,12,12]]$ BB codes at realistic physical error rates. For the distance $5$ surface code at physical noise $p=0.003$, our neural post-selector achieves logical error suppression comparable to the leading complementary gap method without running the decoder.
\end{abstract}

\maketitle

\section{Introduction}
Current quantum computers have hundreds of noisy qubits, and, at present, do not achieve quantum advantage over conventional means for operationally useful applications.
Quantum error correction (QEC) is necessary to enable large-scale fault-tolerant quantum computation, and improving the efficiency of QEC protocols is central to efforts to build industrially- and scientifically-relevant quantum computers.
Recent years have seen rapid progress on both the theoretical side of QEC -- such as the introduction of novel and more efficient quantum codes~\cite{Breuckmann_2021,panteleev2022asymptoticallygoodquantumlocally,Bravyi2024} and decoders~\cite{Panteleev_2021,LSD,muller2025improvedbeliefpropagationsufficient} -- and on the experimental side, where small scale experimental demonstrations of codes and logical operations have been achieved~\cite{Bluvstein_2023,Google,Gupta_2024}.

An important technique used to achieve low logical error rates (LER) in experiments is \emph{post-selection}: identifying which shots are most likely to result in a logical error -- instances in which the decoder fails -- and rejecting them.
While the LER decreases exponentially with code distance below the threshold, the low error rates necessary for practical applications require large code distances and, consequently, a substantial qubit overhead.
Post-selection on the other hand allows one to achieve LER comparable to those of larger codes with a fraction of the qubits and is thus expected to be a critical tool to enable useful fault-tolerant quantum computing in the near future~\cite{Smith_2024}.
The price to pay for this reduction in qubits is a lower \emph{yield} (the fraction of shots accepted), which increases runtime.

\begin{figure}[t]
    \centering
    \includegraphics[width=\linewidth]{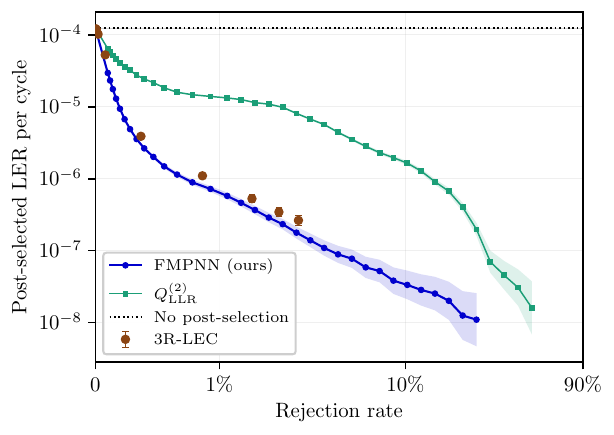}
    \caption{Neural post-selection on the $[[72,12,6]]$ bivariate bicycle code under uniform depolarising circuit-level noise at a physical error rate $p=0.001$, contrasted with popular benchmarks for quantum low-density parity-check code post-selection: $Q_{\text{LLR}}^{(2)}$~\cite{lee2026efficientpostselectiongeneralquantum} and the three-round logical-error criterion (3R-LEC)~\cite{xie2026simpleefficientgenericpostselection}. LER per-cycle is defined as $1-(1-\text{LER})^\frac{1}{T}$, for $T$ cycles.}
    \label{fig:teaser}
\end{figure}

The trade-off between LER and yield is favourable whenever discarding a shot is cheap relative to incurring a logical fault, for instance in magic state preparation~\cite{PRXQuantum.5.010302} and distillation~\cite{Smith_2024}, scalable preparation of resource states~\cite{staples2026scalablepostselectionquantumresources}, and logical error mitigation, where decoder or syndrome information is used to either reject shots~\cite{dinca2026errormitigationlogicalcircuits,zhou2025errormitigationfaulttolerantquantum}, or to condition the mitigation~\cite{aharonov2025syndromeawaremitigationlogical}.

\begin{table*}[t]
    \centering
    \small
    \setlength{\tabcolsep}{4pt}
    \providecommand{\asym}[3]{\ensuremath{#1^{+#2}_{-#3}}}

    \caption{\textbf{Per-cycle logical error rate suppression at 10\% rejection.} Higher is better, errors are 95\% intervals. 3R-LEC ranges span the nearest
    measured points, with rejection percentages in parentheses.
    $^{*}$: cutoff; $\dagger$: fewer than five accepted failures.}
    \label{tab:main-results}

    \resizebox{\textwidth}{!}{%
    \begin{tabular}{llccccc}
        \toprule
        Code & $p$ & Neural & Density & $Q_{\text{LLR}}^{(2)}$ & Gap & 3R-LEC (Rej.\ \%) \\
        \midrule
        Surface $[[25,1,5]]$
        & 0.001 & $\dagger$ & \asym{7.4}{0.3}{0.3}
        & \asym{151.5}{24.0}{20.7} & $\dagger$ & $19.9$ ($1.54^{*}$) \\
        & 0.003 & \asym{96.5}{3.4}{3.3} & 2.0
        & 9.6 & \asym{104.2}{3.9}{3.7} & 9.4--11.6 (5.54--10.15) \\

        \addlinespace
        Surface $[[49,1,7]]$
        & 0.001 & $\dagger$ & \asym{4.4}{0.4}{0.4}
        & \asym{301.0}{224.9}{128.6} & $\dagger$ & $42.4$ ($0.77^{*}$) \\
        & 0.003 & \asym{70.5}{3.3}{3.2} & 1.7
        & 7.1 & \asym{463.3}{58.3}{51.7} & 15.7--17.4 (4.95--10.89) \\

        \addlinespace
        Surface $[[81,1,9]]$
        & 0.001 & $\dagger$ & \asym{3.4}{0.9}{0.7}
        & $\dagger$ & $\dagger$ & $\dagger$ ($0.56^{*}$) \\
        & 0.003 & \asym{20.5}{0.8}{0.8} & 1.6
        & \asym{6.7}{0.1}{0.1} & \asym{2096.1}{1022.8}{687.3}
        & 22.4--26.4 (4.73--12.80) \\

        \addlinespace
        BB $[[72,12,6]]$
        & 0.001 & \asym{3716.6}{2158.6}{1365.4} & 5.3
        & \asym{70.6}{4.6}{4.3} & n/a & $469.8$ ($2.66^{*}$) \\

        \addlinespace
        BB $[[144,12,12]]$
        & 0.003 & \asym{738.2}{171.1}{138.9} & 1.6
        & \asym{70.0}{4.6}{4.3} & n/a & 75.2--381.0 (6.92--20.30) \\
        \bottomrule
    \end{tabular}%
    }
\end{table*}

In this work, we introduce two post-selection models: the Message Passing Neural Network (MPNN) for surface codes, and the factor MPNN (FMPNN) for high-rate quantum low-density parity-check (qLDPC) codes, which satisfy the following three desiderata:

\begin{enumerate}
    \item \textbf{Generality:} the architectures apply across code families and decoders. Their input is the syndrome (surface codes) or the syndrome together with the decoder's own belief-propagation (BP) posteriors (qLDPC codes).
    \item \textbf{Specialisable:} each model is trained on \texttt{(input, label)} pairs obtained from running the decoder of interest, and can be fine-tuned to desired noise models and error rates. Since it operates on the detector error model (DEM), it applies in principle to any circuit-level procedure.
    \item \textbf{Single-staged:} post-selection involves a single forward pass per shot on a GPU, with no repeated decoder calls.
\end{enumerate}

Our work treats post-selection as a binary classification task: each shot is labelled with $Y=1$ if the decoder succeeds and $Y=0$ if it fails. The architecture is decoder-agnostic, with the caveat that qLDPC codes require a BP-based decoder. Labelled data can be generated cheaply and on demand by simulation, as we do here, or directly by hardware, for example by a quantum memory experiment that checks whether the decoder preserves a known logical state.

Our contributions are:
\begin{itemize}
    \item \textbf{A learned graph neural post-selector for qLDPC codes.} We evaluate the FMPNN on the $[[72,12,6]]$ and $[[144,12,12]]$ bivariate bicycle (BB) codes~\cite{Bravyi2024} under circuit-level noise. The FMPNN achieves the best LER suppression (Table~\ref{tab:main-results}) vs yield trade-offs at all rejection rates amongst methods that do not execute the decoder multiple times. It further matches or outperforms the three-round logical-error criterion (3R-LEC) variant of argument reweighting~\cite{xie2026simpleefficientgenericpostselection}, which requires 3 passes of the decoder.

    \item \textbf{A specialised model for surface codes.} The MPNN requires only the syndrome at inference, meaning it can be executed before decoding. We evaluate it on the rotated surface codes at distances $5,7,9$ under circuit-level noise. The MPNN achieves the best LER suppression at all rejection rates amongst methods which do not execute the decoder multiple times. At distance $5$ and $p=3\times10^{-3}$, it achieves similar LER suppression at all rejection rates to the complementary gap~\cite{hutter2014efficient}, the leading post-selection strategy for surface codes, with higher throughput.

    \item \textbf{A cost analysis.} We give the graph-size dependence of the neural post-selection models (Section~\ref{sec:timingcost}) and measure the per-shot processing times empirically, comparing against other common baselines on the same hardware.
\end{itemize}

Recently, several works have proposed post-selection criteria for QEC.
These approaches fall into three broad families.

\textit{The first reuses decoder confidence}: posterior estimates, BP log-likelihood ratios (LLRs), or cluster-level soft information from localized statistics decoding (LSD)~\cite{meister2024efficientsoftoutputdecoderssurface,dinca2026errormitigationlogicalcircuits,zhou2025errormitigationfaulttolerantquantum, lee2026efficientpostselectiongeneralquantum,kishi2026efficientsoftoutputdecodingextracluster}. These criteria can be inexpensive to extract, but depend on the decoder used.

\textit{The second family compares repeated decoder executions.} The complementary gap compares correction weights across opposite logical outcomes~\cite{hutter2014efficient,PRXQuantum.5.010302, gidney2023yokedsurfacecodes, gidney2024magic}; argument reweighting modifies the error model and checks agreement between successive logical predictions~\cite{xie2026simpleefficientgenericpostselection}. Forced gap applies separate logical constraints for $k$ logical observables using $k+1$ decoder calls~\cite{wills2026forcedgappostselectionquantum}.

\textit{The third family learns confidence from data.}
Neural decoders can support post-selection through per-logical confidence estimates~\cite{gu2026scalableneuraldecoderspractical}, graph neural network (GNN)~\cite{PhysRevResearch.7.023181} logits used as complementary gap proxies~\cite{dentelski2026neural}, and transformer confidence outputs~\cite{seip2026machinelearningoptimalquantum}. Haug et al.\ distinguish between low and high noise syndromes~\cite{haug2026machinelearnedsyndromepostselectionreliable}. Bhambay et al.\ use partial syndrome histories for adaptive aborting during syndrome extraction~\cite{bhambay2026adaptiveabortingschemesquantum}. In contrast, our models predict the success of an existing decoder.

Post-selection has also been studied through statistical-mechanical analyses of abort and conditional logical thresholds~\cite{English_2025}, and large-deviation arguments for scalable logical-error suppression~\cite{chen2025scalableaccuracygainspostselection}. These approaches provide theoretical insights into the scalability of post-selection.

The remainder of this paper is organised as follows.
Section~\ref{sec:Preliminaries} introduces the basics of quantum error correcting codes, decoders, and post-selection in more detail. Section~\ref{sec:Method} describes the MPNN and FMPNN architectures and training scheme. Section~\ref{sec:Results} presents our results of post-selected memory experiments on surface codes and BB codes for a variety of settings. Section~\ref{sec:Conclusions, Outlook and Future Work} concludes and outlines directions for future work.

\section{Preliminaries}
\label{sec:Preliminaries}

\subsection{Stabiliser codes and syndrome extraction}
\label{sec:QECprelims}
A $[[n,k,d]]$ stabiliser code encodes $k$ logical qubits into $n$ physical qubits and corrects up to $\lfloor \frac{d-1}{2} \rfloor$ errors. The codespace is defined as the simultaneous $+1$ eigenspace of a set of commuting Pauli operators, the elements of which are called stabiliser generators. Errors are diagnosed by repeatedly measuring these generators; a single round of measurement produces a syndrome bit per generator, across $T$ rounds of syndrome extraction we obtain the syndrome vector $s \in \{0,1\}^{mT}$, where $m$ is the number of generators. For circuit-level noise, it is convenient to work with \textit{detectors}: parities of stabiliser measurements that evaluate to $0$ in the absence of noise. The detector vector $\sigma \in \{0,1\}^r$ for $r$ detectors is the input to all decoders and post-selectors considered here. The relationship between fault mechanisms and detectors is encoded by the DEM, equivalent to a Tanner graph $\mathcal{G} = \{\mathcal{V}_{\text{fault}} \bigcup \mathcal{V}_{\text{det}}, E \}$ in which each fault mechanism in $\mathcal{V}_{\text{fault}}$ is connected to the detectors in $\mathcal{V}_{\text{det}}$ it triggers. Writing $F:= |\mathcal{V}_{\text{fault}}|$, the parity check matrix is the adjacency matrix $H \in \mathbb{F}_2^{r \times F}$ of the Tanner graph, where rows correspond to detectors and columns to fault mechanisms. The logical action matrix (or logical observable matrix) $A\in \mathbb{F}_2^{k\times F}$ contains one row per logical observable: $A_{if} = 1$ when fault $f$ toggles observable $i$. For a binary fault vector $e$, the detector syndrome is $\sigma =He$ and the logical outcome is $L=Ae$. A correction $c$ consistent with $\sigma$ is successful when $Ac = Ae$.

\subsection{Code families}
Our experiments focus on rotated surface code and BB code memories under circuit-level noise.

\textbf{Rotated surface codes}~\cite{Dennis,KITAEV20032,rotated1,rotated2}. A planar $[[d^2,1,d]]$ topological code with weight-4 $X$- and $Z$-type stabilisers in the bulk and weight-2 stabilisers at the boundary. We study $d=5,7$ and $9$ instances.

\textbf{Bivariate bicycle codes}~\cite{Bravyi2024,TourdeGross}. A family of qLDPC codes constructed from two commuting polynomials over $\mathbb{F}_2[x,y]/(x^l -1, y^m-1 )$. They achieve higher encoding rates and competitive distances to that of the surface code, but require non-local connectivity. We study the $[[72,12,6]]$ and $[[144,12,12]]$ instances.

\subsection{Decoders}
We benchmark using the following two decoders:

\subsubsection{Minimum Weight Perfect Matching}

The surface code admits a decomposition to a \textit{graph-like} DEM, where each fault results in the flip of at most two detectors. This property allows it to be decoded using Minimum Weight Perfect Matching (MWPM). We use PyMatching~\cite{Higgott2025sparseblossom} as our MWPM implementation.

\subsubsection{Belief propagation and Localized Statistics Decoding}

In general, qLDPC codes do not have graph-like DEMs and hence we cannot use MWPM; instead, it is common to use a two-stage general purpose decoder, consisting of BP~\cite{pearl2022reverend} and some post-processing step, in our case, LSD~\cite{LSD}. We provide the full implementation details and parameters for the decoders in Appendix~\ref{App:Decoderconfig}.

\subsection{Post-selection in quantum error correction}
We describe post-selection in the following way. Denote $x$ for the information available to a post-selector. Given a confidence score $s(x) \in \mathbb{R}$ for each shot, and threshold $\tau \in \mathbb{R}$, the rule accepts a shot iff $s(x) \geq \tau$. Larger scores indicate greater confidence in decoder success. The performance of the post-selector is characterised by two quantities:

\begin{equation}
    \label{eq:yield}
    \text{Yield}(\tau) = \mathbb{P}(s(x) \geq \tau),
\end{equation}
\begin{equation}
    \label{eq:PSLER}
    \text{LER}(\tau) = \mathbb{P}(Ac(\sigma) \neq Ae \mid s(x) \geq \tau),
\end{equation}
where $A$ is the logical action matrix of Section~\ref{sec:QECprelims}. In practice the yield is the fraction of retained shots and the conditional LER is defined when the yield is non-zero. A good post-selector significantly suppresses LER at yields acceptable to the application.

\begin{figure*}[ht]
    \centering
    \includegraphics[width =\linewidth]{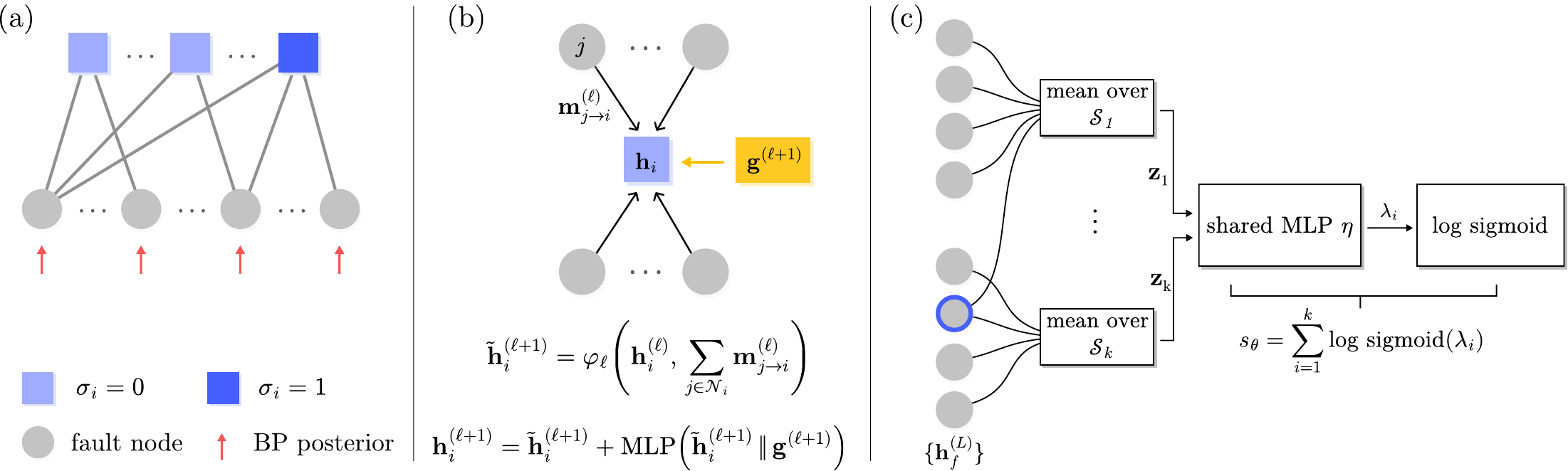}
    \caption{Factor-graph MPNN (FMPNN). (a) Detector and fault nodes carry syndrome, circuit and belief-propagation features. (b) Message passing combines local and global context. (c) Fault embeddings are pooled over their logical supports; overlapping supports can share nodes (highlighted). A shared multilayer perceptron produces per-observable logits, which are used to produce the final confidence score.}
    \label{fig:MPNNQLDPC}
\end{figure*}
\textbf{Existing post-selection strategies.}

\begin{itemize}
    \item \textit{Complementary gap~\cite{hutter2014efficient}:} For the surface code, the complementary gap re-runs the decoder once under the constraint that the returned correction lies in the complementary logical class, and uses the log-likelihood difference between the constrained and unconstrained corrections, the gap, as the confidence score. A small gap indicates the decoder cannot confidently distinguish the two outcomes and such shots are rejected. This construction underpins recent surface-code protocols for magic state preparation~\cite{PRXQuantum.5.010302} and magic state cultivation~\cite{gidney2024magic}. The method does not extend directly to high-rate qLDPC codes as they are not matchable in general and also have $4^k$ logical classes, rendering a naive enumeration of the joint outcomes prohibitively expensive.
    \item \textit{Forced gap~\cite{wills2026forcedgappostselectionquantum}:} After an initial decode, this method forces the complementary outcome of each of the $k$ logical observables and uses the smallest increase in correction weight as confidence. It requires $k+1$ decoder executions. Using Relay-BP~\cite{muller2025improvedbeliefpropagationsufficient}, it has been shown to achieve efficient gap-based post-selection for high-rate qLDPC codes. As the forced-gap method has been established for Relay-BP and our FMPNN model was trained for use in conjunction with BP+LSD, we do not directly compare the two approaches.
    
    \item \textit{Argument reweighting~\cite{xie2026simpleefficientgenericpostselection}:}
    A related construction for qLDPC codes by re-decoding a desired number of times under a modified error model which is re-weighted to suppress the previous correction(s), accepting the shot only when the subsequent decodes agree with the first. Re-decoding can be done an arbitrary number of times for higher performance at the cost of increased runtime.

    \item \textit{Neural Decoder logit-based post-selection:} The Cascade neural decoder~\cite{gu2026scalableneuraldecoderspractical} provides confidence estimates for post-selection. The GNN decoder of Lange et al.~\cite{PhysRevResearch.7.023181} supplies the logit used to perform post-selection~\cite{dentelski2026neural}; in both cases the score is specific to the neural decoder that produces it.

    \item \textit{Other Neural post-selection techniques:} Bhambay et  al.~\cite{bhambay2026adaptiveabortingschemesquantum} introduce neural predictors for adaptively aborting  shots during syndrome extraction. Their scheme uses a convolutional neural network to estimate the probability of a logical failure from partial syndrome history, aborting and restarting when it exceeds a desired threshold. Haug et al.~\cite{haug2026machinelearnedsyndromepostselectionreliable} train a classifier to distinguish syndrome samples generated at low and high physical error rates, rejecting shots whose syndrome is predicted to be associated with a high physical error rate with probability higher than a specified threshold.

    \item \textit{Correction weight~\cite{lee2026efficientpostselectiongeneralquantum}:} For a returned correction $c$ and DEM prior probabilities $p_j$, define:

    \[
    w(c) = \sum_j c_j\log \frac{1-p_j}{p_j}.
    \]
    A high-weight correction reflects an improbable error and is treated as less trustworthy, shots whose correction weight exceeds a given threshold are rejected. This procedure is essentially free once the decoder has returned a correction, however it tends to require very low yields to achieve meaningful LER suppression.

    \item \textit{Cluster-based confidence scores:} Meister et al.~\cite{meister2024efficientsoftoutputdecoderssurface} extract soft-information from clusters grown by MWPM or union-find decoding for the surface code. Setting edge weights to zero within these clusters they use the minimum remaining weight of a path connecting inequivalent boundaries as a confidence score, where a small score indicates little additional error weight is  needed to produce a logical fault, and such shots are rejected. Lee et al.~\cite{lee2026efficientpostselectiongeneralquantum} develop confidence scores from the clusters produced by LSD~\cite{LSD}. Assigning each cluster the total LLR weight of the faults it contains, the statistic $Q_{\text{LLR}}^{(2)}$ is the Euclidean norm of these per-cluster weights divided by the total  weight of all mechanisms: shots with larger values are less trustworthy and rejected.

    \item \textit{Detector density:} A simple baseline of $s_{\text{density}}(\sigma) = 1- \frac{1}{r}\sum_i\sigma_i$, where shots are discarded once the fraction of detectors which fired passes a specified cut-off. Its usage is motivated by the desirable property of running before decoder execution. It has also been shown to exhibit a  statistical mechanical abort phase transition for post-selection of surface codes in the code-capacity setting~\cite{English_2025}.
\end{itemize}

Our experiments compare our models with that of detector density, correction weight, the LSD cluster statistics, argument reweighting, and the complementary gap for surface codes.

\textbf{Related uses of machine learning in QEC.} Neural approaches to QEC also address decoding, such as topological-code decoders~\cite{Torlai}, neural BP~\cite{NBP}, equivariant decoders~\cite{egorov2023endequivariantneuraldecoder}, the transformer-based AlphaQubit and AlphaQubit 2~\cite{Alpha,Alpha2}, and the state-space Mamba decoder of~\cite{mamba}. On the encoding side, machine learning has also been used for the discovery and design of QEC codes~\cite{Olle2024,BO,meyer2026learning,cruz2026evolutionary,sella2026designing,qian2026multi, yan2026omniqec}. Neural pre-decoders for surface~\cite{chamberland2026fastaccurateaibasedpredecoders} and colour codes~\cite{olle2026fastaccurateaibasedpredecoders} pass simplified residual syndromes to a downstream decoder to improve its reliability; unlike post-selection, no shots are rejected.

\section{The MPNN Architecture}
\label{sec:Method}
\begin{figure*}[t]
    \centering
    \includegraphics[width = \linewidth]{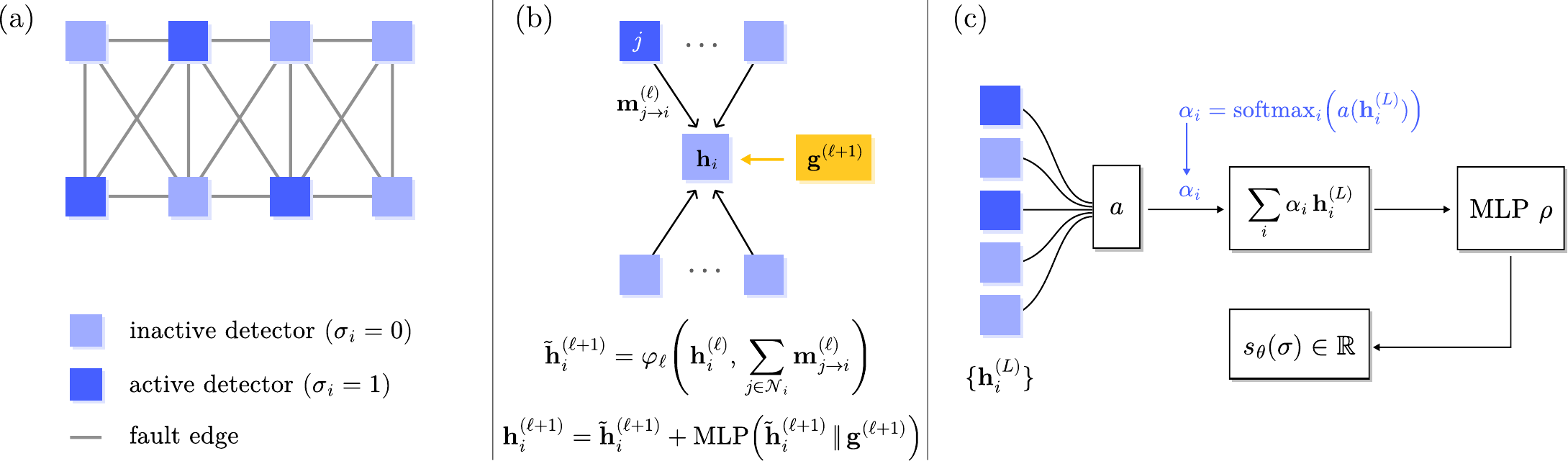}
    \caption{Surface-code MPNN. (a) Detector graph labelled by the
    syndrome. (b) Message passing combines local and global context. (c) Attention pooling and a multilayer perceptron produce a confidence score from the final embeddings of the detectors.}
    \label{fig:MPNNSC}
\end{figure*}

\subsection{Problem statement}
Let $\sigma \in \{0,1\}^r$ denote the detector vector of a single shot and $Y = \mathbbm{1}[Ac(\sigma) = Ae]$ indicate successful decoding. For a fixed circuit and decoder, let $x = (\sigma, b_{BP})$ for the BB-model input and $x= \sigma$ for the surface-model input, where $b_{BP}$ denotes the BP features defined in Section~\ref{sec:bpinput}. We train GNNs to predict $Y$.

We first describe the factor-graph model for general qLDPC codes, then specialise the model to graph-like decoding for surface codes.

\subsection{Factor MPNN for qLDPC codes}
\label{sec:FMPNNIntro}
For general qLDPC codes, a single fault can trigger three or more detectors, and as such, it need not admit a graph-like decomposition. The FMPNN operates on the full bipartite DEM graph (Figure~\ref{fig:MPNNQLDPC}), with both detector nodes and fault nodes. We use this model for our evaluations on BB codes.

\subsubsection{Graph and node features}
\label{sec:fmpnnfeats}
Each detector has six features: its syndrome bit, alongside 5 static features: three spacetime coordinates, a boundary indicator and a boundary-fault probability. The fault node represents each fault mechanism and is connected to each detector it flips, each fault node has three static features: its prior probability $p_f$, the LLR $\log\frac{1-p_f}{p_f}$, and its degree, in addition to the per-shot BP features of Section~\ref{sec:bpinput}. Let $\mathbf{x}_i^{\text{det}} \in \mathbb{R}^6, \mathbf{x}_f^{\text{fault}} \in \mathbb{R}^6$ be the respective detector and fault node features. Two learned input maps initialise the hidden embeddings:

\begin{equation}
    \label{eq:embeddingdet}
    \mathbf{h}_i^{(0)} = \varphi_{\text{det}}(\mathbf{x}_i^{\text{det}}) \in \mathbb{R}^h,
\end{equation}
\begin{equation}
    \label{eq:embeddingfault}
    \mathbf{h}_f^{(0)} = \varphi_{\text{fault}}(\mathbf{x}_f^{\text{fault}}) \in \mathbb{R}^h,
\end{equation}
where both maps consist of a learned linear layer followed by ReLU. $h$ is the hidden dimension held fixed across message-passing layers.

\subsubsection{Belief-propagation input features}
\label{sec:bpinput}
For qLDPC codes, it is common to use BP-based decoders, which offer rich soft-information which can be used as an input to the GNN in addition to the syndrome. We additionally give the following three BP features to each fault node: the posterior LLR of the fault, the difference between the posterior and prior, and BP's hard decision.

\subsubsection{Message-passing layers}
\label{sec:mplayers}
The network applies $L$ rounds of local message passing between neighbouring detector and fault nodes:

\begin{equation}
    \mathbf{m}_{j \rightarrow i}^{(\ell)} := \psi_\ell(\mathbf{h}_i^{(\ell)},\mathbf{h}_j^{(\ell)}),
\end{equation}
\begin{equation}
    \tilde{\mathbf{h}}_{i}^{(\ell+1)} := \varphi_\ell(\mathbf{h}_i^{(\ell)},\sum_{j \in \mathcal{N}_i} \mathbf{m}^{(\ell)}_{j \rightarrow i}),
\end{equation}
where $\mathcal{N}_i$ is the neighbourhood of node $i$. $\psi_\ell$ is a small messaging function which constructs the information sent from neighbour $j$ to node $i$, it concatenates their embeddings and applies a two-layer multilayer perceptron (MLP):

\begin{equation}
    \label{eq:Psidef}
    \psi_\ell(\mathbf{u},\mathbf{v}) = W_2\text{ReLU}(W_1[\mathbf{u||v}] + \mathbf{b}_1) + \mathbf{b}_2,
\end{equation}
where $\mathbf{u}$, $\mathbf{v}$ are two node embeddings. Here, weight matrices $W_1, W_2$ and biases $\mathbf{b}_1, \mathbf{b}_2$ are learned; each message is of dimension $h$.

For a node embedding $\mathbf{u}$ and summed incoming messages $\mathbf{m}$, the update function $\varphi_\ell$ is an MLP given by

\begin{equation}
    \mathbf{r} = U_2\text{ReLU}(U_1[\mathbf{u||m}] + \mathbf{c}_1) + \mathbf{c}_2,
\end{equation}

\begin{equation}
    \label{eq:phidef}
    \varphi_\ell(\mathbf{u},\mathbf{m}) = \mathbf{u} + \text{ReLU}(\text{BN}_\ell(\mathbf{r})).
\end{equation}
Here, $U_1,U_2, \mathbf{c}_1, \mathbf{c}_2$ are the MLP's learned weights and biases, and $\text{BN}_\ell$ is the batch normalisation at layer $\ell$. Each layer has separate messages and update parameters which are shared across both node types.

The receptive field of the local message-passing grows with $L$. FMPNN edges have no features and just describe connectivity. Each layer is followed by a global context update described in Section~\ref{sec:global}, and final fault embeddings are pooled over logical supports as described in Section~\ref{sec:supportpool}.

\subsubsection{Global context update}
\label{sec:global}
After each local update, we mean-pool the intermediate embeddings over all detector and fault nodes:

\begin{equation}
    \mathbf{g}^{(\ell + 1)} := \frac{1}{|\mathcal{V}|} \sum_{v \in \mathcal{V}}\tilde{\mathbf{h}}_v^{(\ell +1)},
\end{equation}
with $|\mathcal{V}| = r+F$, and $\tilde{\mathbf{h}}_v^{(\ell +1)}$ is used to denote the intermediate embedding after the $\ell$-th message-passing update but before the global context update. The global update is broadcast back to every node, through a shared MLP and residual connection,

\begin{equation}
    \mathbf{h}_{v}^{(\ell +1)} = \tilde{\mathbf{h}}_v^{(\ell+1)} + \mathrm{MLP}_{\mathrm{global}}\bigl(\tilde{\mathbf{h}}_v^{(\ell+1)} || \ \mathbf{g}^{(\ell+1)}\bigr),
\end{equation}
where $\tilde{\mathbf{h}}_v^{(\ell+1)} || \ \mathbf{g}^{(\ell+1)}$ is the vector formed from concatenation of $\tilde{\mathbf{h}}_v^{(\ell+1)}$ and $\mathbf{g}^{(\ell+1)}$. Empirically, models with this component outperformed those without it.

\subsubsection{Per-logical support pooling}
\label{sec:supportpool}
A memory experiment tracking $k$ logical observables has $k$ individual success labels and $2^k$ joint outcomes. We use one prediction head per observable. Row $i$ of the logical action matrix $A$ defines the support of logical observable $i$, the set of fault mechanisms that flip it: $\mathcal{S}_i = \{f: A_{if} =1\}.$

\begin{equation}
    \mathbf{z}_i = \frac{1}{|\mathcal{S}_i|}\sum_{f \in \mathcal{S}_i} \mathbf{h}_f^{(L)}, \qquad i = 1,\dots,k,
\end{equation}
and a shared small 2-layer MLP $\eta$ maps each pooled vector to a per-observable success logit $\lambda_i = \eta(\mathbf{z}_i)$. Detector embeddings enter the readout only indirectly, through their influence on fault embeddings during message passing. The acceptance score derives from the product of the per-observable success probabilities which we compute as

\begin{equation}
    s_\theta(x)= \sum_{i=1}^k \log \text{sigmoid}(\lambda_i),
\end{equation}
where we take the logarithmic form for numerical stability, larger scores favour acceptance. Each sigmoid output $p_i = \text{sigmoid}(\lambda_i)$ is trained to estimate the marginal success probability for logical observable $i$.
The per-logical support idea was inspired by the Cascade neural decoder~\cite{gu2026scalableneuraldecoderspractical}, where data qubits were pooled in a logical support.

\subsection{Surface-code specialisation}
\label{sec:MPNNIntro}
For the surface-code memory circuits we consider here, the DEM can be decomposed into graph-like components each affecting at most two detectors~\cite{Gidney2021stimfaststabilizer,Higgott2025sparseblossom}. This permits MWPM decoding and a simpler detector graph representation for our MPNN.

\subsubsection{Detector graph and inputs}

Each surface code detector carries six features: its syndrome bit, and five static features: its spacetime coordinates \texttt{(x,y,round)}, a boundary indicator for detectors affected by single-detector components, and a boundary-fault probability. We use Stim's~\cite{Gidney2021stimfaststabilizer} graph-like DEM decomposition, which represents each fault by correlated components affecting at most two detectors each. A two-detector component contributes an edge between those detectors, while a single-detector component sets that detector's boundary indicator. We encode the single-detector contribution as node features rather than introducing boundary nodes. Their combined probability is given by $p_B = \frac{1-\prod_j (1-2p_j)}{2}$, where $j$ indexes independent DEM faults contributing a single-detector component at that detector with probability $p_j$.

For each edge connecting detectors $u,v$, contributions from two-detector components on that pair are combined using the same rule, writing the resulting probability as $p_{uv}$, the edge additionally carries the following two features:
\begin{equation}
    \label{eq:Edgefeat}
    e_{uv} =\left(p_{uv},\,\log\frac{1-p_{uv}}{p_{uv}}\right),
\end{equation}
which are the prior fault probability and its LLR. This graph encodes contributions from all components but does not explicitly encode their correlations. Figure~\ref{fig:MPNNSC} depicts this representation. All features except the syndrome are fixed by the circuit and noise model.

\subsubsection{Message passing and readout}
\label{sec:MPNNMP&RO}
A learned linear layer followed by ReLU embeds each detector into the hidden space of dimension $h$ similarly as in Section~\ref{sec:fmpnnfeats}. Message passing now includes the edge features:

\begin{equation}
    \label{eq:SCMP}
    \mathbf{m}_{j \rightarrow i}^{(\ell)} = \psi_\ell^{\text{SC}}(\mathbf{h}_i^{(\ell)}, \mathbf{h}_j^{(\ell)}, e_{ij}),
\end{equation}
where the message function $\psi_\ell^{\text{SC}}$ follows the same setup as in Eq.~\ref{eq:Psidef}, but now the edge features are additionally concatenated with the two detector embeddings, the update function $\varphi_\ell$ remains of the same form as in Eq.~\ref{eq:phidef}. Parameters are learned separately for each layer.

Each layer then applies the global context update of Section~\ref{sec:global} over detector nodes. After $L$ layers of message passing we use attention pooling followed by an MLP readout $\rho$ to produce the final score:
\begin{equation}
    \label{eq:SCscore}
    s_\theta(\sigma) = \rho\left(\sum_{i=1}^r\alpha_i\mathbf{h}_i^{(L)}\right),
\end{equation}
where $\alpha_i = \text{softmax}_i\left(a(\mathbf{h}_i^{(L)})\right)$, for a small gating MLP $a$. Both $a$ and $\rho$ are two-layer MLPs with separate learned weights and biases:

\begin{align*}
    &a(\mathbf{z}) = \mathbf{w}_a^\mathsf{T}\text{ReLU}(W_a\mathbf{z} + \mathbf{b}_a) + c_a,\\
    &\rho(\mathbf{z}) = \mathbf{w}_\rho^\mathsf{T}\text{ReLU}(W_\rho\mathbf{z} + \mathbf{b}_\rho) + c_\rho.
\end{align*}

\begin{figure*}[!t]
    \centering
    \includegraphics[width=\linewidth]{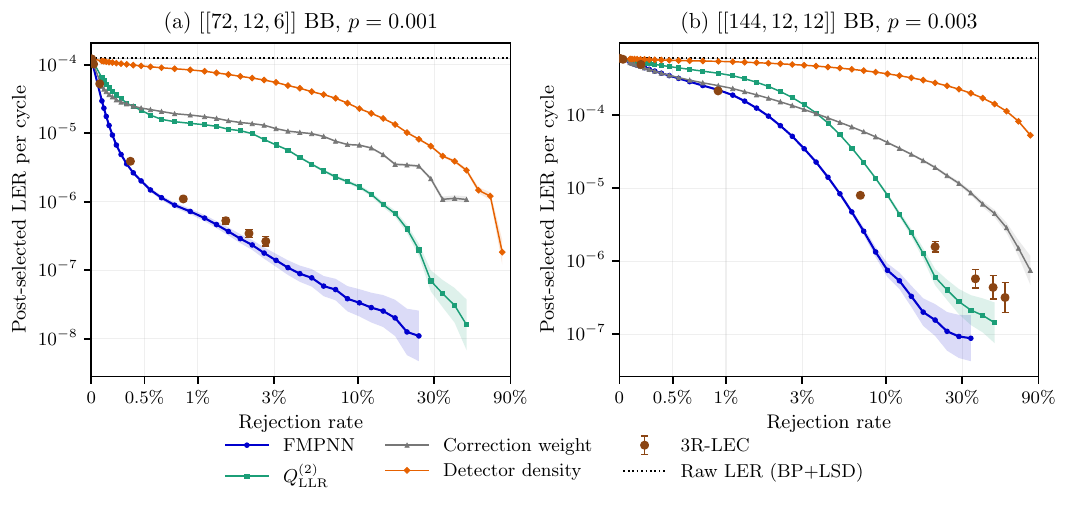}
    \caption{Post-selected logical error rate against rejection rate for the (a) $[[72,12,6]]$ and (b) $[[144,12,12]]$ bivariate bicycle codes under uniform depolarising circuit-level noise at physical error rates $p=0.001$ and $p=0.003$ respectively.}
    \label{fig:ResultsBB}
\end{figure*}

\subsection{Training objective}
\label{sec:training}
For BB codes, the per-observable success labels are $Y_i = \mathbbm{1}[(Ac(\sigma))_i = (Ae)_i]$, where $c(\sigma)$ is the decoder's returned correction. The shot-level label is $Y=\prod_{i=1}^k Y_i$. We minimise binary cross-entropy (BCE) for these individual labels. For surface code training, we use weighted BCE for the shot-level label $Y$. Written in terms of the success logit $z$, the BCE is
\begin{equation}
    \label{BCE}
    \ell_{\text{BCE}}(y,z) = y\,\text{softplus}(-z) + (1-y)\,\text{softplus}(z).
\end{equation}

The primary objectives are
\begin{align}
    &\mathcal{L}_{\text{BB}} = \mathbb{E}\left[\frac{1}{k}\sum_{i=1}^k \ell_{\text{BCE}}(Y_i,\lambda_i)\right], \\
    & \mathcal{L}_{\text{SC}} = \mathbb{E}\left[w(Y) \ell_{\text{BCE}}(Y,s_\theta)\right].
\end{align}
Here $w(Y)$ is a weighting term to address the highly imbalanced classes, which we use for sampling at higher error rates. At lower error rates, we use importance sampling. Training is specific to the chosen decoder: in our case, BP+LSD for BB codes and MWPM for surface codes. Appendix~\ref{app:training} gives further details on our training procedure.

\subsection{Post-selection pipeline}
The acceptance threshold $\tau$ is calibrated for the desired LER (or yield). For high-rate qLDPC codes, BP supplies features to the FMPNN which then scores the shots, and the post-processing (LSD in our case) is executed only for accepted shots. For surface codes, given a syndrome, only one forward pass of the MPNN is required to produce the score, and accepted shots are then passed onto MWPM for decoding. The neural networks do not propose a correction or require repeated decoder passes, unlike alternative methods. Section~\ref{sec:timingcost} reports the costs, with further empirical results in Appendix~\ref{app:additionalresults}.

\subsection{Computational cost}
\label{sec:timingcost}
Let $V$ and $E$ be the number of nodes and undirected edges. We count operations per shot for $L$ layers of message passing with hidden dimension $h$. The matrix-vector multiplication costs $O(h^2)$, so evaluating the message MLP on each edge and the update MLP on each node costs $O(L(V+E)h^2)$. One can avoid repeating the same matrix products for neighbouring edges resulting in an optimised implementation reducing this cost to
\begin{equation}
    \label{eq:MPcost}
    O\left(L(Vh^2 +Eh)\right),
\end{equation}
details of this optimisation can be found in Appendix~\ref{App:Cost}.

At each layer, computing the global mean costs $O(Vh)$, and combining it with every node through the MLP costs $O(Vh^2)$. The final attention readout for the surface code MPNN computes one scalar per detector, with an $h \rightarrow h/2 \rightarrow 1$ MLP, costing $O(Vh^2)$. Softmax costs $O(V)$, the weighted sum costs $O(Vh)$, and the output MLP costs $O(h^2)$. Thus these contributions are within the bound of Eq.~\ref{eq:MPcost}.

For the high-rate qLDPC codes, the logical support pooling of Section~\ref{sec:supportpool} pools $F$ fault embeddings over $k$ logical supports, costing $O(Fhk)$, applying the output MLP to each pooled embedding adds $O(kh^2)$. The total scoring cost of the FMPNN is then

\begin{equation}
    \label{eq:BBcost}
    O\left((LE+kF)h + (LV+k)h^2\right).
\end{equation}

The dominant factor which affects the cost of deploying the models is therefore the size of the hidden dimension, which we keep moderately small at $h=32$ and $h=64$ for BB and surface codes respectively. Using a simple global mean instead of logical support pooling would reduce the pooling cost to $O(Vh)$, however we retain the present readouts for their enhanced performance.

\section{Results}
\label{sec:Results}

\begin{figure*}[!t]
    \includegraphics[width = \linewidth]{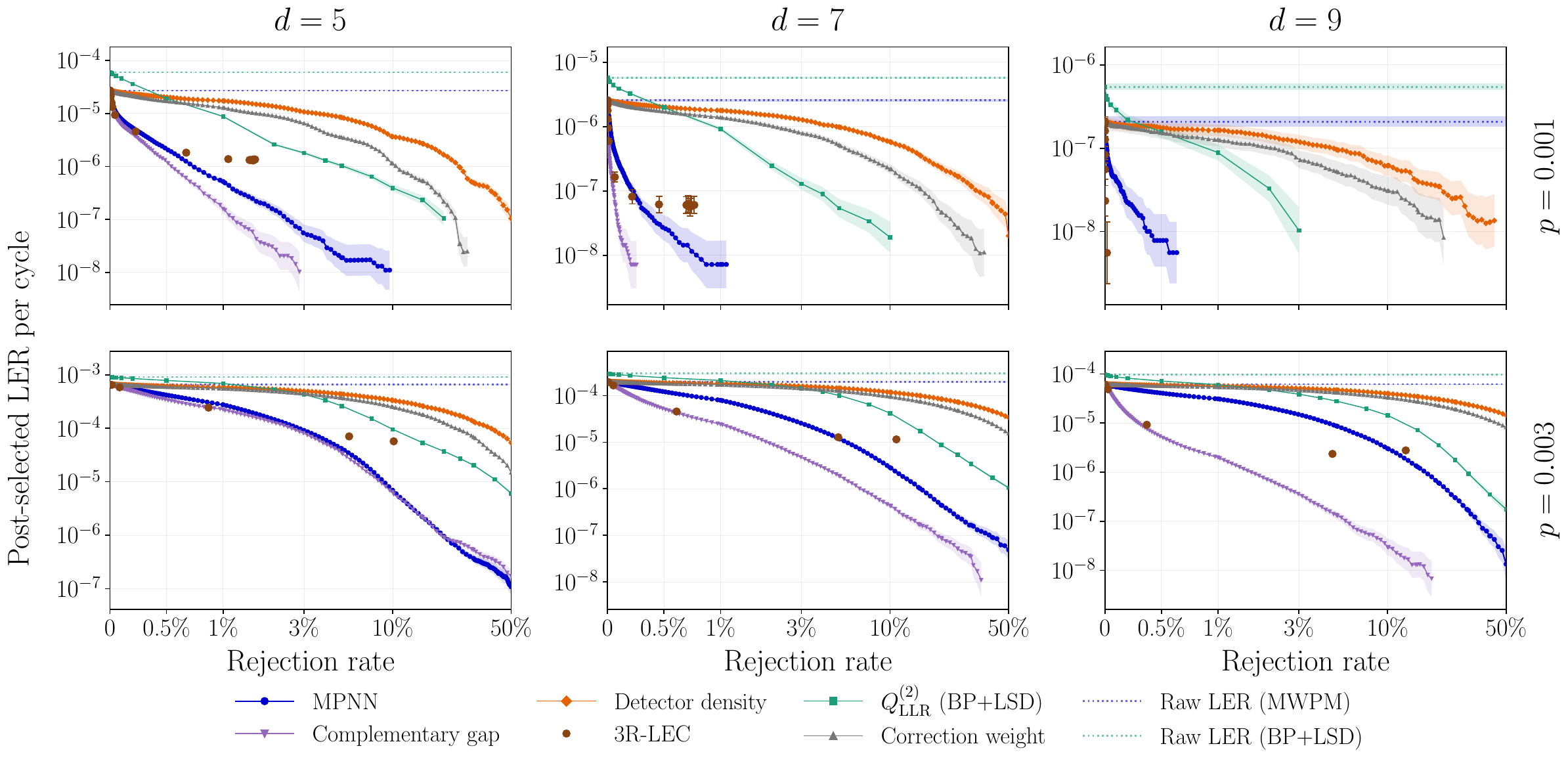}
    \caption{Post-selected logical error rate against rejection rate for the rotated surface codes under uniform depolarising circuit-level noise at distances $5,7,9$ and physical noise $p=0.001$ and $p=0.003$. Each $(d,p)$ configuration uses a separately fine-tuned MPNN; see Appendix~\ref{sec:surface-training-details}.}
    \label{fig:ResultsSC}
\end{figure*}

We illustrate the performance of the FMPNN and MPNN on BB code and surface code memories through plotting the conditional post-selected LER per cycle against the rejection rate, the fraction of shots discarded (lower-left is better). For all codes we use $T=d$ syndrome extraction cycles and circuit level depolarising noise as described in Appendix~\ref{App:NoiseModel}. Shaded areas denote $95\%$ confidence. We additionally provide empirical timings comparing the throughput of the various methods.

\subsection{Bivariate bicycle codes}
We evaluate the $[[72,12,6]]$ and $[[144,12,12]]$ BB codes of~\cite{Bravyi2024} under uniform depolarising circuit-level noise at physical error rates $p=10^{-3}$ and $3\times 10^{-3}$ respectively. Figure~\ref{fig:ResultsBB} shows the trade-off between LER reduction and yield, where we achieve the best performance amongst compared methods for both codes. For the 72-qubit code, rejecting just $1\%$ of shots reduces the LER from $1.24\times10^{-4}$ to $6.41\times10^{-7}$; at $10\%$ rejection, the LER is reduced by a factor of approximately 3700, accepting only 18 of 74,290 failures, catching 99.976\%. For the 144-qubit code, at a higher noise regime, $10\%$ rejection reduces the LER from $6.02\times10^{-4}$ to $8.15\times10^{-7}$, a factor of 738. The corresponding factors at $10\%$ rejection for $Q_{\text{LLR}}^{(2)}$ are 70.6 and 70. The FMPNN also compares favourably with 3R-LEC, especially once a larger rejection budget is available due to the known saturation of LER reduction as rejection increases~\cite{xie2026simpleefficientgenericpostselection}. The post-selected LERs of 3R-LEC at $6.92\%$ ($b=1.1$) and $20.3\%$ ($b=1.2$) rejection are $8\times 10^{-6}$ and $1.58 \times 10^{-6}$ respectively. Using an NVIDIA A100 GPU and single-threaded CPU, at $10\%$ rejection, our FMPNN pipeline takes 1.09\,ms and 11.40\,ms per attempted shot for the 72- and 144-qubit BB codes respectively, compared with 0.97\,ms and 11.86\,ms for the single threaded CPU $Q_{\text{LLR}}^{(2)}$. Here, GPU inference with the FMPNN makes up 0.052\,ms and 0.584\,ms of these totals, with the remainder being CPU decoding work (see Table~\ref{tab:processing-times}). The same two 3R-LEC points take $55.64$\,ms and $104.48$\,ms per attempted shot.

\subsection{Surface codes}
We study circuit-level $Z$-basis memory of $d=5,7,9$ rotated surface codes under uniform depolarising circuit-level noise at physical error rates $p \in \{0.001,0.003\}$. Figure~\ref{fig:ResultsSC} compares the MPNN post-selection performance with other available baselines, where we achieve the best LER suppression amongst methods only requiring one decoder run. Here, $Q_\text{LLR}^{(2)}$ uses BP+LSD to obtain its cluster-level information, while all others use MWPM. At $p= 3 \times 10^{-3}$, and $10\%$ rejection, the MPNN suppresses the LER by factors of $96.5,70.5, \text{and } 20.5$ for $d=5,7,9$. The BP+LSD cluster-score factors relative to its own raw LER are $9.56, 7.13, \text{and } 6.71$. Table~\ref{tab:main-results} summarises these values alongside a selection of other methods.

At $d=5, p =3 \times 10^{-3}$, the MPNN achieves similar performance to that of the complementary gap, while the gap retains a significant advantage at larger distances. At $p=10^{-3}$, the complementary gap, 3R-LEC and MPNN all achieve significant LER suppression within less than $1\%$ rejection. At $p=3\times10^{-3}$ and $10\%$ rejection, the complete post-selection and decoding pipeline takes $4.91$\,\textmu s, $18.42$\,\textmu s, and $48.71$\,\textmu s at $d=5,7,9$ respectively, which is $2.8\times$ to $3.3\times$ faster than the single-threaded complementary gap implementation (Table~\ref{tab:processing-times}).

\section{Conclusions, Outlook and Future Work}
\label{sec:Conclusions, Outlook and Future Work}
We introduce a GNN which operates on the native DEM of a quantum error correcting code, and approaches the logical error reduction performance of state-of-the-art techniques, in a single GPU forward pass, without requiring multiple decoder runs. Because training needs only shots labelled by the deployed decoder, from simulation or hardware, the approach carries over to other codes and decoders.

\textbf{Rare-event sampling:} As we pursue larger codes and/or lower physical error rates, the regime of vanishingly small LERs introduces the well-known rare event sampling problem where it becomes prohibitively expensive to collect enough failing shots to train the MPNN. This is an outstanding challenge across QEC more broadly, and an active area of research~\cite{BravyiRare,SubsetSampling,beverland2025failfasttechniquesprobe, mayer2025rareeventsimulationquantum}. Progress there would allow MPNN training at lower LERs.

\textbf{Application to Fault-tolerant procedures:}
Training FMPNN models for usage beyond memory in fault-tolerant protocols which already heavily rely on post-selection such as distillation~\cite{MSD1} or cultivation~\cite{gidney2024magic} is an interesting next step.

\textbf{Throughput optimisation:}
In order to avoid the accumulation of errors during a QEC cycle, one would like to decode at a throughput of at least one syndrome-extraction round per microsecond, thus requiring our post-selector to also fit within these timescales. Promising work has been done on optimising GNNs for field-programmable gate array (FPGA) application at these timescales for high energy physics~\cite{HongxiangGNNFPGA}, which motivates further work in this area for post-selection.

\textit{Model availability:}
The MPNN and FMPNN weights and codebase will shortly be available at 
\href{https://github.com/conorcarty/MPNN4QEC}{\texttt{https://github.com/conorcarty/MPNN4QEC}}.

\textit{Acknowledgements:}
We acknowledge computational resources and support provided by the Imperial College Research Computing Service (\url{http://doi.org/10.14469/hpc/2232}), the Imperial College London Department of Computing GPU Cluster, and the Edinburgh Compute and Data Facility (\url{https://information-services.ed.ac.uk/research-support/research-computing/ecdf}). JR is funded by an EPSRC Quantum Career Acceleration Fellowship (grant code: UKRI1224). TN is supported by the Engineering and Physical Sciences Research Council (grant number EP/W524384/1), the University of Edinburgh and Quantinuum. This work was supported in part by a grant of access to OpenAI models through the ChatGPT for Academic Researchers program.

\textit{Declaration of Generative AI usage: }
The core ideas of the paper were conceived by the authors. CC acknowledges the use of Claude Series 4-5 models (Sonnet, Opus, Fable), and GPT 5.6 Sol, GPT 6 Astra to assist with code development and debugging, performance optimisations, preparation of figures and tables, and manuscript drafting, editing, and review. CC reviewed and edited the resulting material and takes responsibility for its inclusion in this work.

\FloatBarrier
\def\bibpreamble{\enlargethispage{\baselineskip}}
\bibliography{bib}

\appendix
\pagebreak
\onecolumngrid
\raggedbottom

\section{Supplementary Material}

\subsection{Additional Results}
\label{app:additionalresults}
In this section, we provide additional results detailing the LER suppression and timings of our models compared with other popular post-selection techniques. Tables~\ref{tab:LER1}, \ref{tab:LER5}, \ref{tab:LER20} show the LER suppression achieved by each method at fixed rejection rates $1\%,5\%,$ and $20\%$. Table~\ref{tab:processing-times} provides further information on the execution times for each post-selection strategy.

\subsubsection{Logical Error Suppression}
\begin{table}[H]
    \centering
    \small
    \setlength{\tabcolsep}{4pt}

    \caption{\textbf{Per-cycle logical error rate suppression at 1\% rejection.}
    Errors are 95\% intervals. 3R-LEC ranges span the nearest
    measured points, with rejection percentages in parentheses.
    $^{*}$: cutoff; $\dagger$: fewer than five accepted failures.}
    \label{tab:LER1}

    \resizebox{\textwidth}{!}{%
    \begin{tabular}{llccccc}
        \toprule
        Code & $p$ & Neural & Density & $Q_{\text{LLR}}^{(2)}$ & Gap & 3R-LEC (Rej.\ \%) \\
        \midrule
        Surface $[[25,1,5]]$
        & 0.001 & \asym{52.0}{6.7}{5.9} & 1.6
        & \asym{6.8}{0.2}{0.2} & \asym{171.2}{42.3}{33.9}
        & 14.8--19.6 (0.67--1.07) \\
        & 0.003 & 2.4 & 1.1
        & 1.3 & 3.0 & 2.7--9.4 (0.87--5.54) \\

        \addlinespace
        Surface $[[49,1,7]]$
        & 0.001 & \asym{355.2}{476.0}{203.1} & 1.4
        & \asym{6.2}{0.5}{0.4} & $\dagger$ & $42.4$ ($0.77^{*}$) \\
        & 0.003 & 2.5 & 1.1
        & 1.4 & \asym{8.2}{0.1}{0.1} & 4.4--15.7 (0.61--4.95) \\

        \addlinespace
        Surface $[[81,1,9]]$
        & 0.001 & $\dagger$ & \asym{1.3}{0.1}{0.1}
        & \asym{6.2}{1.4}{1.1} & $\dagger$ & $\dagger$ ($0.56^{*}$) \\
        & 0.003 & 2.0 & 1.1
        & 1.6 & \asym{31.1}{1.5}{1.4} & 6.7--26.4 (0.37--4.73) \\

        \addlinespace
        BB $[[72,12,6]]$
        & 0.001 & \asym{193.1}{20.3}{18.4} & 1.5
        & \asym{9.1}{0.2}{0.2} & n/a & 112.8--234.7 (0.86--1.50) \\

        \addlinespace
        BB $[[144,12,12]]$
        & 0.003 & 2.9 & 1.1
        & 1.7 & n/a & 2.8--75.2 (0.93--6.92) \\
        \bottomrule
    \end{tabular}
    }
\end{table}

\begin{table}[H]
    \centering
    \small
    \setlength{\tabcolsep}{4pt}

    \caption{\textbf{Per-cycle logical error rate suppression at 5\% rejection.}
    Errors are  95\% intervals. 3R-LEC ranges span the nearest measured points, with rejection percentages in parentheses.
    $^{*}$: cutoff; $\dagger$: fewer than five accepted failures.}
    \label{tab:LER5}

    \resizebox{\textwidth}{!}{%
    \begin{tabular}{llccccc}
        \toprule
        Code & $p$ & Neural & Density & $Q_{\text{LLR}}^{(2)}$ & Gap & 3R-LEC (Rej.\ \%) \\
        \midrule
        Surface $[[25,1,5]]$
        & 0.001 & \asym{1425.0}{1283.2}{675.0} & 3.2
        & \asym{57.6}{5.3}{4.8} & $\dagger$ & $19.9$ ($1.54^{*}$) \\
        & 0.003 & \asym{16.0}{0.2}{0.2} & 1.5
        & 3.5 & \asym{19.3}{0.3}{0.3} & 2.7--9.4 (0.87--5.54) \\

        \addlinespace
        Surface $[[49,1,7]]$
        & 0.001 & $\dagger$ & \asym{2.6}{0.2}{0.2}
        & \asym{105.9}{40.5}{29.2} & $\dagger$ & $42.4$ ($0.77^{*}$) \\
        & 0.003 & \asym{16.8}{0.4}{0.4} & 1.4
        & 2.9 & \asym{106.4}{6.0}{5.7} & 15.7--17.4 (4.95--10.89) \\

        \addlinespace
        Surface $[[81,1,9]]$
        & 0.001 & $\dagger$ & \asym{2.2}{0.4}{0.3}
        & $\dagger$ & $\dagger$ & $\dagger$ ($0.56^{*}$) \\
        & 0.003 & \asym{7.1}{0.2}{0.1} & 1.3
        & 3.4 & \asym{465.8}{93.6}{77.9} & 22.4--26.4 (4.73--12.80) \\

        \addlinespace
        BB $[[72,12,6]]$
        & 0.001 & \asym{1535.1}{512.2}{384.0} & 3.0
        & \asym{33.6}{1.4}{1.4} & n/a & $469.8$ ($2.66^{*}$) \\

        \addlinespace
        BB $[[144,12,12]]$
        & 0.003 & \asym{65.0}{4.0}{3.8} & 1.4
        & \asym{10.3}{0.2}{0.2} & n/a & 2.8--75.2 (0.93--6.92) \\
        \bottomrule
    \end{tabular}
    }
\end{table}

\begin{table}[H]
    \centering
    \small
    \setlength{\tabcolsep}{4pt}

    \caption{\textbf{Per-cycle logical error rate suppression at 20\% rejection.}
    Errors are  95\% intervals. 3R-LEC ranges span the nearest measured points, with rejection percentages in parentheses.
    $^{*}$: cutoff; $\dagger$: fewer than five accepted failures.}
    \label{tab:LER20}

    \resizebox{\textwidth}{!}{%
    \begin{tabular}{llccccc}
        \toprule
        Code & $p$ & Neural & Density & $Q_{\text{LLR}}^{(2)}$ & Gap & 3R-LEC (Rej.\ \%) \\
        \midrule
        Surface $[[25,1,5]]$
        & 0.001 & $\dagger$ & \asym{16.3}{1.3}{1.2}
        & \asym{567.5}{199.4}{147.5} & $\dagger$ & $19.9$ ($1.54^{*}$) \\
        & 0.003 & \asym{685.3}{71.6}{64.8} & 3.1
        & \asym{24.6}{0.4}{0.4} & \asym{720.6}{77.3}{69.8}
        & $11.6$ ($10.15^{*}$) \\

        \addlinespace
        Surface $[[49,1,7]]$
        & 0.001 & $\dagger$ & \asym{11.2}{2.0}{1.7}
        & $\dagger$ & $\dagger$ & $42.4$ ($0.77^{*}$) \\
        & 0.003 & \asym{451.5}{59.6}{52.7} & 2.4
        & \asym{34.3}{1.0}{0.9} & \asym{2294.5}{738.5}{558.7}
        & $17.4$ ($10.89^{*}$) \\

        \addlinespace
        Surface $[[81,1,9]]$
        & 0.001 & $\dagger$ & \asym{6.0}{2.7}{1.8}
        & $\dagger$ & $\dagger$ & $\dagger$ ($0.56^{*}$) \\
        & 0.003 & \asym{102.8}{10.1}{9.2} & 2.0
        & \asym{27.1}{1.1}{1.0} & $\dagger$ & $22.4$ ($12.80^{*}$) \\

        \addlinespace
        BB $[[72,12,6]]$
        & 0.001 & \asym{8495.0}{9041.7}{4379.7}
        & \asym{11.8}{0.3}{0.3}
        & \asym{298.8}{44.4}{38.7} & n/a & $469.8$ ($2.66^{*}$) \\

        \addlinespace
        BB $[[144,12,12]]$
        & 0.003 & \asym{3609.2}{2253.9}{1387.3} & 2.2
        & \asym{874.9}{238.0}{187.1} & n/a & 75.2--381.0 (6.92--20.30) \\
        \bottomrule
    \end{tabular}%
    }
\end{table}

\subsubsection{Empirical Time Costs}

\begin{table}[H]
    \caption{Complete processing time in \textmu s per attempted shot, including required decoding. All methods except 3R-LEC target $10\%$ rejection and report means $\pm$ $95\%$ confidence intervals. 3R-LEC spans the prescribed sweep as in Appendix~\ref{app:baseline-implementations}. Neural uses an A100 plus one AMD EPYC 7702P thread; other baselines use one such CPU thread.}
    \label{tab:processing-times}
    \centering\small
    \setlength{\tabcolsep}{4pt}
    \begin{tabular}{@{}llrrrrr@{}}
        \toprule
        Code & $p$ & Neural & Density & $Q^{(2)}_{\rm LLR}$ & Gap & \shortstack{3R-LEC\\$b$ sweep}\\
        \midrule
        Surface $d=5$ & $10^{-3}$ & $4.14\pm0.01$ & $0.426\pm0.004$ & $601\pm3$ & $12.05\pm0.03$ & 990--1,000\\
        Surface $d=5$ & $3\times10^{-3}$ & $4.91\pm0.01$ & $1.194\pm0.003$ & $931\pm1$ & $16.42\pm0.01$ & 1,602--1,611\\
        Surface $d=7$ & $10^{-3}$ & $15.64\pm0.03$ & $1.32\pm0.01$ & $2{,}398\pm5$ & $41.75\pm0.06$ & 4,760--4,810\\
        Surface $d=7$ & $3\times10^{-3}$ & $18.42\pm0.03$ & $4.20\pm0.01$ & $4{,}250\pm14$ & $56.42\pm0.04$ & 5,274--5,334\\
        Surface $d=9$ & $10^{-3}$ & $41.80\pm0.07$ & $3.11\pm0.01$ & $6{,}995\pm17$ & $99.43\pm0.10$ & 11,118--11,240\\
        Surface $d=9$ & $3\times10^{-3}$ & $48.71\pm0.07$ & $9.93\pm0.03$ & $11{,}738\pm8$ & $138.11\pm0.09$ & 11,348--11,436\\
        \midrule
        BB72 & $10^{-3}$ & $1{,}092\pm71$ & $765\pm7$ & $965\pm27$ & n/a & 2,757--4,317\\
        BB144 & $3\times10^{-3}$ & $11{,}403\pm468$ & $9{,}561\pm140$ & $11{,}864\pm1{,}322$ & n/a & 35,514--292,206\\
        \bottomrule
    \end{tabular}
\end{table}

\subsubsection{Optimised Message Passing}
\label{App:Cost}
The optimisation in Section~\ref{sec:timingcost} reuses matrix products within each layer of the MPNN and FMPNN architectures. Splitting the message MLP's first weight matrix gives $W_1[\mathbf{h}_i||\mathbf{h}_j] = W_r\mathbf{h}_i + W_s \mathbf{h}_j$. Both products are computed once per node per layer and reused, costing $O(Vh^2)$. MPNN messages also include two edge features which contribute $O(h)$ per edge and may be cached in the case of a fixed graph. Combining terms and applying ReLU costs $O(Eh)$. Writing $\mathbf{u}_{ji}$ as the vector after ReLU, the final message matrix multiplication can be applied after summation:
\[
    \sum_{j\in \mathcal{N}_i} \mathbf{m}_{j \rightarrow i} = W_2 \sum_{j\in \mathcal{N}_i}\mathbf{u}_{ji} + |\mathcal{N}_i|\mathbf{b}_2.
\]
So, $W_2$ is applied once per node. Message aggregation costs $O(Eh)$ and node updates cost $O(Vh^2)$, giving Eq.~\ref{eq:MPcost} for $L$ layers.

\subsection{Circuit noise description}
\label{App:NoiseModel}

\begin{table}[H]
    \caption{Circuit-level noise model. All surface-code experiments use Stim's generated \texttt{surface\_code:rotated\_memory\_z} circuit at code distance $d$ with $d$ rounds of syndrome extraction, under uniform depolarising
    noise: a single physical error rate $p$ is applied to all four channels below.}
    \label{tab:noise-model}
    \centering

    \begin{tabular}{@{}llc@{}}
        \toprule
        Stim parameter & Inserted channel (location) & Rate \\
        \midrule
        \texttt{after\_clifford\_depolarization}      & \texttt{DEPOLARIZE1}/\texttt{DEPOLARIZE2} after each 1-/2-qubit gate & $p$ \\
        \texttt{before\_round\_data\_depolarization}  & \texttt{DEPOLARIZE1} on data qubits, start of each round            & $p$ \\
        \texttt{before\_measure\_flip\_probability}   & \texttt{X\_ERROR} before each measurement                           & $p$ \\
        \texttt{after\_reset\_flip\_probability}      & \texttt{X\_ERROR} after each reset                                  & $p$ \\
        \bottomrule
    \end{tabular}
\end{table}

\begin{table}[H]
    \caption{Circuit-level noise model for the bivariate
    bicycle experiments, using the unmodified circuit
    builder of~\cite{gong2024toward}, distributed through
    \cite{lee2026efficientpostselectiongeneralquantum}.
    The $[[72,12,6]]$ and $[[144,12,12]]$ codes use
    $Z$-basis memory circuits with six and twelve rounds
    of syndrome extraction, respectively. All listed
    channels use the same physical error rate $p$.
    }
    \label{tab:bb-noise-model}
    \centering
    \begin{tabular}{@{}p{0.22\textwidth}p{0.68\textwidth}c@{}}
        \toprule
        Operation/location & Inserted channel & Rate \\
        \midrule
        Two-qubit gates
        & \texttt{DEPOLARIZE2} after each CNOT
        & $p$ \\

        Data-qubit idles
        & \texttt{DEPOLARIZE1} on the left data block during
        the first CNOT layer of every round, and on the
        right data block at the start of each round after
        the first
        & $p$ \\

        Initial preparation
        & \texttt{X\_ERROR} after each initial $Z$-basis
        reset of data and ancilla qubits
        & $p$ \\

        Ancilla reinitialisation
        & \texttt{X\_ERROR} on $Z$-check ancillas and
        \texttt{Z\_ERROR} on $X$-check ancillas,
        at the start of each subsequent round
        & $p$ \\

        Check measurements
        & \texttt{X\_ERROR} before $Z$-check measurements
        (\texttt{MR}) and \texttt{Z\_ERROR} before
        $X$-check measurements (\texttt{MRX})
        & $p$ \\
        \bottomrule
    \end{tabular}
\end{table}

\subsection{Decoder configurations}
\label{App:Decoderconfig}
For the rotated surface codes, we use MWPM, implemented using the sparse-blossom algorithm in \texttt{PyMatching}~\cite{Higgott2025sparseblossom}. The matching graph is constructed with \texttt{decompose\_errors=True}. We use standard matching without correlated matching enabled. For BB codes (and rotated surface codes post-selected using the clustering approach of~\cite{lee2026efficientpostselectiongeneralquantum}), we use BP+LSD and retain the BP parameters of the original LSD configuration~\cite{LSD}: min-sum BP with a maximum of 30 iterations, with a scaling factor of $\alpha = 0.625$, and a parallel update schedule. We use order 0 LSD. We implement BP+LSD using the soft-output implementation released with~\cite{lee2026efficientpostselectiongeneralquantum}, built upon the \texttt{ldpc} package~\cite{LDPC}. All qLDPC training labels and every baseline we report are computed from this configuration.

\subsection{Baseline implementations}
\label{app:baseline-implementations}
Table~\ref{tab:baseline-sources} identifies the software used for each baseline. The decoders are sourced from \href{https://github.com/oscarhiggott/PyMatching}{\texttt{PyMatching}}~2.3.1~\cite{Higgott2025sparseblossom} and \href{https://github.com/quantumgizmos/ldpc}{\texttt{ldpc}}~2.4.1~\cite{LDPC,LSD}. We obtain soft outputs through Lee et al.'s \href{https://github.com/seokhyung-lee/ldpc-post-selection}{\texttt{ldpc-post-selection}}~0.1.1~\cite{lee2026efficientpostselectiongeneralquantum}. 

\begin{table}[H]
    \caption{Sources of the implementations for compared baselines.}
    \label{tab:baseline-sources}
    \centering\small
    \renewcommand{\arraystretch}{1.15}
    \begin{tabularx}{\linewidth}{@{}>{\raggedright\arraybackslash}p{0.23\linewidth}X@{}}
        \toprule
        Baseline & Implementation and source\\
        \midrule
        Detector density &  The \texttt{detector\_density} output of Lee et al.'s~\cite{lee2026efficientpostselectiongeneralquantum} \href{https://github.com/seokhyung-lee/ldpc-post-selection}{\texttt{ldpc-post-selection}} code. \\
        Correction weight & The \texttt{pred\_llr} output of Lee et al.'s \texttt{SoftOutputsMatchingDecoder} for surface codes and \texttt{SoftOutputsBpLsdDecoder} for BB codes. \\
        Complementary gap & Lee et al.'s \texttt{SoftOutputsMatchingDecoder} is used for the curves produced in Figure~\ref{fig:ResultsSC}. Timing analyses for the gap are done using Gidney et al.'s~\cite{gidney2024magic} \href{https://github.com/Strilanc/magic-state-cultivation}{\texttt{magic-state-cultivation}} repo's \texttt{PymatchingGapSampler}. \\
        $Q^{(2)}_{\rm LLR}$ & Lee et al.'s \texttt{SoftOutputsBpLsdDecoder}.\\
        3R-LEC &  Xie et al.'s~\cite{xie2026simpleefficientgenericpostselection} \href{https://github.com/qHaipengXie/Argument-Reweighting}{\texttt{Argument-Reweighting}} code: \texttt{MWPM.py} for surface codes and \texttt{BPLSDPOST.py} for BB codes.\\
        \bottomrule
    \end{tabularx}
\end{table}

\paragraph{3R-LEC parameters.}
Surface code runs use MWPM; BB runs use the BP+LSD settings as in Appendix~\ref{App:Decoderconfig}. We implement 3R-LEC using the suggested parameters of Ref.~\cite{xie2026simpleefficientgenericpostselection}, with $b=1+\delta$:
\par\noindent{\small%
\begin{tabular}{@{}l@{\quad}l@{}}
    SC, $p=10^{-3}$: & $\delta\in\{10^{-12}, 10^{-6}, 10^{-3}, 0.01, 0.02, 0.04, 0.06, 0.1, 0.2, 0.3, 0.4, 0.5, 0.6, 0.7, 0.8, 0.9, 1\}$ \\
    SC, $p=3\times10^{-3}$: & $\delta\in\{10^{-3}, 0.01, 0.1, 0.3, 0.5\}$ \\
    BB: & $\delta\in\{10^{-15}, 10^{-8}, 10^{-4}, 10^{-3}, 0.01, 0.1, 0.2, 0.3, 0.4, 0.5\}$
\end{tabular}}

\subsection{Training details}
\label{app:training}

\subsubsection{Bivariate bicycle models}
\label{app:bb-training-details}
\begin{table}[H]
    \caption{Architectures and inputs of the two final FMPNN models. The five-output auxiliary head adds $613$ parameters and is unused at inference. Belief-propagation inputs are required at inference as well as during training. The number of message-passing layers and hidden dimension were chosen after non-exhaustive ablations and further optimisations could improve model performance.}
    \label{tab:bb-final-architectures}
    \centering
    \small
    \renewcommand{\arraystretch}{1.13}
    \begin{tabularx}{\linewidth}{@{}>{\raggedright\arraybackslash}p{0.35\linewidth}XX@{}}
        \toprule
        Setting & BB72 & BB144 \\
        \midrule
        Code parameters & $[[72,12,6]]$ & $[[144,12,12]]$ \\
        Reported physical error rate & $10^{-3}$ & $3\times10^{-3}$ \\
        Syndrome-extraction rounds & $6$ & $12$ \\
        Detector nodes & $252$ & $936$ \\
        Fault-mechanism nodes & $2592$ & $10368$ \\
        Message-passing layers & $4$ & $12$ \\
        Hidden dimension & $32$ & $32$ \\
        Logical output heads & $12$ & $12$ \\
        Model parameters at inference & $39\,137$ & $115\,425$ \\
        \midrule
        Graph & \multicolumn{2}{p{0.61\linewidth}@{}}{Bipartite detector-mechanism graph from the undecomposed circuit DEM.} \\
        Detector node features (6) & \multicolumn{2}{p{0.61\linewidth}@{}}{Syndrome bit; three detector coordinates; boundary indicator; boundary-fault probability.} \\
        Fault-mechanism node features (6) & \multicolumn{2}{p{0.61\linewidth}@{}}{Prior fault probability, prior LLR, degree, BP posterior LLR, BP hard decision, and posterior-minus-prior LLR.} \\
        BP feature normalisation & \multicolumn{2}{p{0.61\linewidth}@{}}{Posterior LLR and posterior-minus-prior LLR are each clipped to $[-60,60]$ and divided by $20$.} \\
        Message passing & \multicolumn{2}{p{0.61\linewidth}@{}}{Sum aggregation; ReLU activations, residual connections and batch normalisation; global mean context in every layer.} \\
        Readout & \multicolumn{2}{p{0.61\linewidth}@{}}{Mean pooling over the fault-mechanism support of each logical observable, followed by one shared MLP.} \\
        \bottomrule
    \end{tabularx}
\end{table}

\begin{table}[H]
    \caption{Shared training procedure for the bivariate bicycle models. The $12$ sigmoid outputs are combined into the acceptance score $\sum_{i=1}^{12}\log\text{sigmoid}(\lambda_i)$.}
    \label{tab:bb-training-procedure}
    \centering
    \small
    \renewcommand{\arraystretch}{1.13}
    \begin{tabularx}{\linewidth}{@{}>{\raggedright\arraybackslash}p{0.29\linewidth}>{\raggedright\arraybackslash}X@{}}
        \toprule
        Training setting & Procedure for both BB models \\
        \midrule
        Training data & Online circuit-noise samples at the fixed error rate of each stage.\\
        Labels and BP inputs & Min-sum BP, maximum $30$ iterations, scaling factor $0.625$, parallel updates, followed by order-$0$ LSD. The binary target for logical observable $i$ is $1$ when its decoder prediction is correct. \\
        Primary objective & Unweighted BCE, averaged over the minibatch and the $12$ logical outputs. \\
        Auxiliary objective & Mean-squared error for five standardised targets: correction LLR, $Q_{\text{LLR}}^{(2)}$, $Q_{\text{LLR}}^{(\infty)}$, detector density and correction weight. The total loss is $\mathcal L_{\rm BCE}+\lambda_e\mathcal L_{\rm aux}$. \\
        Auxiliary normalisation & Per-target running means and variances with update coefficient $0.01$; divide centred targets by $\sqrt{v+10^{-6}}$. Statistics are initialised afresh at each trainer invocation, including resumes. \\
        Auxiliary coefficient & Where enabled, $\lambda_e=0.05[1+\cos(\pi(e-1)/(E-1))]$, for stage epoch $e=1,\ldots,E$. Thus $\lambda_e$ decreases from $0.1$ to $0$. The final BB144 stage uses $\lambda_e=0$ throughout. \\
        Optimiser split & Muon for weight matrices whose two dimensions are both at least $16$; Lion for all remaining parameters. \\
        Muon settings & Momentum $0.95$, Nesterov update, five Newton--Schulz iterations; weight decay $10^{-4}$. \\
        Lion settings & Momentum coefficients $(0.9,0.99)$; weight decay $3\times10^{-3}$. \\
        Learning-rate schedule & A fresh cosine schedule in each stage, with minimum-rate parameter $10^{-6}$ for both groups and multiplicative linear warm-up. Stage learning rates and warm-up lengths are listed separately below. \\
        Gradient accumulation & The stage tables give the minibatch size and accumulation factor. Accumulated gradients are clipped to Euclidean norm $1$; a remaining partial accumulation is also updated at the end of each epoch. \\
        Numerical precision & BF16 automatic mixed precision with FP32 model weights; auxiliary standardisation and auxiliary MSE in FP32. \\
        Weight averaging & Exponential moving average (EMA), updated after each optimiser step. Batch-normalisation buffers are copied from the current model. Validation and reported inference use the EMA weights. \\
        Stage initialisation & The first stage starts from random weights. Each subsequent stage loads its predecessor's EMA weights and restarts the optimiser, learning-rate schedule and epoch counter. \\
        \bottomrule
    \end{tabularx}
\end{table}

\begin{table}[H]
    \caption{Training stages in the ancestry of the final BB72 model. There are 250,000 shots per epoch, $b$ is minibatch size, $a$ is the gradient accumulation.}
    \label{tab:bb72-training-stages}
    \centering
    \small
    \setlength{\tabcolsep}{3.3pt}
    \renewcommand{\arraystretch}{1.15}
    \begin{tabular}{@{}lcccccccc@{}}
        \toprule
        Stage & $p$ & Epochs & $b\times a$ & Muon LR & Lion LR & Warm-up & EMA & Seed \\
        \midrule
        Pretraining & $0.003$ & $300$ & $256\times1$ & $2\times10^{-2}$ & $3\times10^{-5}$ & $1000$ & $0.9998$ & $7$ \\
        Initial fine-tuning & $0.001$ & $300$ & $512\times4$ & $2\times10^{-2}$ & $3\times10^{-5}$ & $125$ & $0.998401$ & $7$ \\
        Four fine-tuning continuations & $0.001$ & $300$ & $512\times4$ & $10^{-2}$ & $1.5\times10^{-5}$ & $125$ & $0.998401$ & $8,9,10,12$ \\
        \bottomrule
    \end{tabular}
\end{table}

\begin{table}[H]
    \caption{Training stages of the final BB144 model, with notation as in Table~\ref{tab:bb72-training-stages}. There are 250,000 shots per epoch. }
    \label{tab:bb144-training-stages}
    \centering
    \small
    \setlength{\tabcolsep}{3.3pt}
    \renewcommand{\arraystretch}{1.15}
    \begin{tabular}{@{}lcccccccc@{}}
        \toprule
        Stage & $p$ & Epochs & $b\times a$ & Muon LR & Lion LR & Warm-up & EMA & Seed \\
        \midrule
        Pretraining & $0.003$ & $300$ & $128\times4$ & $2\times10^{-2}$ & $3\times10^{-5}$ & $1000$ & $0.9998$ & $7$ \\
        Continuation & $0.003$ & $150$ & $128\times4$ & $2\times10^{-2}$ & $3\times10^{-5}$ & $1000$ & $0.9998$ & $7$ \\
        Final continuation & $0.003$ & $300$ & $128\times4$ & $1.25\times10^{-3}$ & $2\times10^{-6}$ & $1000$ & $0.9998$ & $20260819$ \\
        \bottomrule
    \end{tabular}
\end{table}

\begin{table}[H]
    \caption{Validation and checkpoint selection for the FMPNN models.}
    \label{tab:bb-checkpoint-selection}
    \centering
    \small
    \renewcommand{\arraystretch}{1.13}
    \begin{tabularx}{\linewidth}{@{}>{\raggedright\arraybackslash}p{0.22\linewidth}>{\raggedright\arraybackslash}X>{\raggedright\arraybackslash}X@{}}
        \toprule
        Selection setting & BB72 & BB144 \\
        \midrule
        Validation population & $10^6$ circuit shots at the stage's physical error rate, fixed within each run. & $10^6$ circuit shots at $p=0.003$, fixed within each run. \\
        Validation seed & $20260617$ & $20260617$ \\
        Validation interval & Every $25$ epochs for pretraining and initial fine-tuning; every $10$ epochs for continuations. & Every $25$ epochs. \\
        Model-selection score & Failure-class average precision (PR-AUC), using EMA predictions. Continuation runs are ranked by the mean of their final five validation values, at epochs $260$--$300$. & Failure-class PR-AUC, using EMA predictions. The final stage retains its best validation checkpoint. \\
        Released weights & Continuation 5, epoch-$300$ EMA; chosen using the fixed validation set. & Final continuation, epoch-$300$ EMA; the saved best-validation checkpoint. \\
        \bottomrule
    \end{tabularx}
\end{table}

\subsubsection{Surface-code models}
\label{sec:surface-training-details}

Tables~\ref{tab:surface-architecture}--\ref{tab:surface-selection} describe the six surface-code models used in the figures. All final fine-tuning and evaluation use the all-component graph construction described in the main text. Five models descend from a mixed-distance parent trained with a different, first-component weighting convention: for a decomposed DEM instruction containing components separated by \texttt{\^{}}, that convention accumulated the instruction probability only into its first component. The final convention includes every graph-like component. 

The $d=5$, $p=10^{-3}$ model instead descends from an independently trained all-component parent. The $d=5$, $p=3\times10^{-3}$ model starts from the earlier distance/noise specialist. The $d=7,9$ models start directly from the mixed-distance parent, extended from three to four and five layers, respectively.

\begin{table}[H]
    \caption{Final surface architectures. The number of message-passing layers and hidden dimension were chosen after non-exhaustive ablations and further optimisations could improve model performance.}
    \label{tab:surface-architecture}
    \centering
    \small
    \renewcommand{\arraystretch}{1.10}
    \begin{tabularx}{\linewidth}{@{}>{\raggedright\arraybackslash}p{0.29\linewidth}XXX@{}}
        \toprule
        Setting & $d=5$ & $d=7$ & $d=9$ \\
        \midrule
        Reported error rates & \multicolumn{3}{l}{$p\in\{10^{-3},3\times10^{-3}\}$} \\
        Syndrome-extraction rounds & $5$ & $7$ & $9$ \\
        Message-passing layers & $3$ & $4$ & $5$ \\
        Hidden dimension & $64$ & $64$ & $64$ \\
        Model parameters & $117\,570$ & $155\,202$ & $192\,834$ \\
        \midrule
        Graph & \multicolumn{3}{p{0.67\linewidth}@{}}{Detector graph from Stim's decomposed circuit DEM.} \\
        Node features (6) & \multicolumn{3}{p{0.67\linewidth}@{}}{Syndrome bit; three detector coordinates; boundary indicator; boundary-fault probability.} \\
        Edge features (2) & \multicolumn{3}{p{0.67\linewidth}@{}}{Fault probability $p_{uv}$ and $\log[(1-p_{uv})/p_{uv}]$.} \\
        Message passing & \multicolumn{3}{p{0.67\linewidth}@{}}{Sum aggregation with two-layer message and update MLPs, ReLU, residual connections and batch normalisation; global mean context in each layer.} \\
        Readout & \multicolumn{3}{p{0.67\linewidth}@{}}{Attention pooling with a $64\to32\to1$ gating MLP and a separate $64\to32\to1$ readout MLP. One success logit ranks shots; larger values favour acceptance.} \\
        \bottomrule
    \end{tabularx}
\end{table}

\begin{table}[H]
    \caption{Final surface fine-tuning: $150$ epochs of $204\,800$ shots.}
    \label{tab:surface-procedure}
    \centering
    \small
    \renewcommand{\arraystretch}{1.10}
    \begin{tabularx}{\linewidth}{@{}>{\raggedright\arraybackslash}p{0.25\linewidth}>{\raggedright\arraybackslash}X@{}}
        \toprule
        Training setting & Final surface fine-tuning procedure \\
        \midrule
        Labels & Binary MWPM success labels, using \texttt{PyMatching} with target-$p$ matching weights; one logical observable per shot. \\
        Optimiser & AdamW, weight decay $10^{-4}$ and default momentum coefficients $(0.9,0.999)$. Each fine-tuning run resets optimiser state and the learning-rate schedule. \\
        Schedule & Full-run cosine decay towards $10^{-6}$, multiplied by a linear warm-up factor; initial-rate parameters and warm-up lengths are in Table~\ref{tab:surface-final-stages}. \\
        Batch updates & One optimiser update per minibatch, with gradient norm clipped to $1$; no gradient accumulation. \\
        Numerics & FP32 weights and operations, with TF32 disabled. Batch-normalisation running moments remain fixed; scale and offset remain trainable. No EMA. \\
        Natural sampling ($p=0.003$) & Online circuit samples. For success label $y\in\{0,1\}$ and logit $z$, the minibatch loss averages $w_+\,y\,\text{softplus}(-z)+(1-y)\,\text{softplus}(z)$. Fixed success weights $w_+$ are $0.00330026$, $0.00142402$ and $0.000547500$ for $d=5,7,9$, respectively. They are failure/success odds estimated from a separate $10^7$-shot population. \\
        Importance sampling ($p=0.001$) & Sample the full undecomposed DEM at proposal $p=0.003$, decode with target-$p$ weights, and minimise importance-weighted BCE. Full-mechanism likelihood ratios are calculated in FP64, clipped at their minibatch $95$th percentile and normalised to sum to one. Both clipping and finite-batch self-normalisation introduce bias. No additional class weighting is applied. \\
        Randomness & Seeds are given below. \\
        \bottomrule
    \end{tabularx}
\end{table}

\begin{table}[H]
    \caption{Final surface runs ($150$ epochs each). Warm-up counts optimiser updates.}
    \label{tab:surface-final-stages}
    \centering
    \small
    \setlength{\tabcolsep}{4pt}
    \renewcommand{\arraystretch}{1.12}
    \begin{tabular}{@{}cccccccc@{}}
        \toprule
        $d$ & $p$ & Initial weights & Batch & Initial LR & Warm-up & Selected epoch & Seed \\
        \midrule
        $5$ & $0.001$ & All-component parent & $256$ & $10^{-4}$ & $1000$ & $140$ & $1648915393$ \\
        $5$ & $0.003$ & Earlier $d=5$ specialist & $512$ & $3\times10^{-5}$ & $800$ & $125$ & $1345664080$ \\
        $7$ & $0.001$ & Mixed parent, $L=4$ & $256$ & $10^{-4}$ & $1600$ & $115$ & $3662741859$ \\
        $7$ & $0.003$ & Mixed parent, $L=4$ & $128$ & $10^{-4}$ & $3200$ & $150$ & $906026237$ \\
        $9$ & $0.001$ & Mixed parent, $L=5$ & $256$ & $10^{-4}$ & $1600$ & $70$ & $74399240$ \\
        $9$ & $0.003$ & Mixed parent, $L=5$ & $128$ & $10^{-4}$ & $3200$ & $145$ & $9051227$ \\
        \bottomrule
    \end{tabular}
\end{table}

\begin{table}[H]
    \caption{Surface checkpoint selection, including epoch zero.}
    \label{tab:surface-selection}
    \centering
    \small
    \renewcommand{\arraystretch}{1.10}
    \begin{tabularx}{\linewidth}{@{}>{\raggedright\arraybackslash}p{0.25\linewidth}>{\raggedright\arraybackslash}X>{\raggedright\arraybackslash}X@{}}
        \toprule
        Selection setting & $d=5$ & $d=7,9$ \\
        \midrule
        Natural validation population at $p=0.001$ & $5\times10^6$ shots & $2\times10^7$ for $d=7$; $10^8$ for $d=9$ \\
        Natural validation population at $p=0.003$ & $10^6$ shots & $10^6$ shots per distance \\
        Rows scored for selection & \multicolumn{2}{p{0.71\linewidth}@{}}{All decoder failures plus $10^5$ controls chosen independently of the labels. Rejection cutoffs are estimated from the controls.} \\
        Selection priority & Fewest retained failures at $5\%$ rejection, then $1\%$, then greatest area under the receiver operating characteristic curve (ROC-AUC). & Fewest retained failures at $1\%$ rejection, then $5\%$, then greatest ROC-AUC. \\
        Validation epochs & $0,1,2,5$, then every $10$ at $p=0.001$ or every $5$ at $p=0.003$, including epoch $150$. & $0,1,2,5$, then every $5$, including epoch $150$. \\
        Final curve evaluation & \multicolumn{2}{p{0.71\linewidth}@{}}{$10^8$ natural shots per distance/noise setting, distinct from the model-selection populations. No importance weights are used for the reported curves.} \\
        \bottomrule
    \end{tabularx}
\end{table}

\paragraph{Parent training and provenance.}
The all-component parent used at $d=5$, $p=10^{-3}$ has three layers and width $64$. It was trained from random weights for $300$ epochs of $204\,800$ shots, with batch size $256$, uniformly over the $25$ cells $d\in\{3,5,7,9,11\}$ and $p\in\{0.001,0.002,0.003,0.004,0.005\}$. It used the numerical and optimiser settings above, initial learning rate $3\times10^{-4}$, $1000$ warm-up updates, fixed success-class weight $1962/498038$, and seed $3385222094$.

The earlier first-component mixed parent also has three layers and width $64$, and was trained with AdamW and cosine warm restarts over $d\in\{3,5,7,9,11\}$ and $p\in[0.001,0.01]$, with seed $7$. The final all-component fine-tuning stages and selected weights are recorded separately in Tables~\ref{tab:surface-final-stages} and~\ref{tab:surface-selection}.

\end{document}